Title: Note on the oblate and prolate deformations in nuclear matter from the viewpoint
   of the quantum-mechanical off-center effect
Author: Mladen Georgiev (Institute of Solid State Physics, Bulgarian Academy of Sciences,
   1784 Sofia, Bulgaria)
Comments: 7 pages including 1 figure, all pdf format
Subj-class: physics

We consider the possibility that a quantum-mechanical off-center effect may be behind the deformed oblate and prolate shapes of nuclei in nuclear physics. In solid state physics, finite off-center displacements result from the mixing of electronic states through their coupling to vibrational (phonon) modes of appropriate symmetries. This is an example of fermion-boson interaction which may materialize in nuclear physics as well in the form of a coupling of nucleons to the π-meson field. We carry our calculations to substantiate the proposal.

1. Foreword

The off-center defects are among the most thrilling objects of interest for solid state physics. Besides their tunneling transitions across the central barrier, they perform barrier-hindered rotations about the central lattice site [1]. The quantum-mechanical entity is conceived smeared about the central barrier and along the reorientational barrier path. As a result of a phonon coupling of appropriate symmetry, its shape may remind the oblate form (symmetry axis and rotational axis perpendicular to each other) or prolate form (symmetry axis and rotational axis parallel to each other) in nuclear physics [2].

Experiments on deformed nuclei have been made casually and details may be found in the reference books [2-4]. We would like to verify whether the nucleus deformations into oblate or prolate forms do not arise from a more fundamental law of nature, that of symmetry breaking due to the mixing of fermionic states through coupling to a boson mode.

It is to be stressed that the spherical shape of a nucleus does not necessarily come at odds with the notion of a deformation. Indeed, the off-center sphere is the result of quantum-mechanical averaging of the reorientational rotation-like motion of off-center ions around the cubic site. Each reorientational position breaks the point-group symmetry (say $O_h$) to a lower symmetry (say $C_{4v}$) and in an idealized case the off-center system if frozen in a $C_{4v}$ position (by virtue of too high a barrier) would clearly demonstrate the symmetry lowering on passing from a cubic configuration to a linear configuration. Now, the point is that on averaging the various segments over their respective angular distribution, a spherical surface results which is the smooth locus of all off-center sites. This highly symmetric sphere around the normal lattice site comes to restore on the average the broken cubic symmetry of that site, though only on the average. It occurs when the $C_{4v}$ parameters are all isotropic in space and assume the same values independent of the coordinate axes.

An illustrative example of a spherical-surface producing system in solid state physics is the $Li^+$ cationic substitution in fcc alkali halides. The on-center $Li^+$ displays the cubic symmetry (e.g. under hydrostatic pressure), but under a $T_{1u}$ odd-mode coupling the cubic symmetry becomes unstable and the impurity goes off-center along any of the six <111> axes.

Experimental techniques, such as paraelectric resonance, paramagnetic resonance, and optical absorption have been employed for data collecting [5].

The deformed spheroid shapes would appear in lower-symmetry cases if, say the z-axis coupling is largely superior to the x- and y-axis ones (prolate shape) or if the x- and y-axis couplings are superior to the z-axis one (oblate shape). Another conceivable case is the octupole (pear shaped) deformation with or without an axis of symmetry [2]. In all the cases mentioned above the experimental analysis of nuclear data would reveal the elliptic semi-axes which in turn would help disclose the kind of nucleus deformation and suggest a mechanism.

## 2. Site-splitting Hamiltonian

Herein we address the off-center defect problem by a method in which the phonon coordinate is regarded as a c-number [6]. The site-splitting adiabatic Hamiltonian (default of the lattice kinetic energy operator) reads

$$H = \sum_{m\alpha} E_{m\alpha} a_{m\alpha}^{\dagger} a_{m\alpha} + \sum_{i\alpha\beta} G_{i\alpha\beta} q_{i\alpha\beta} (a_{m\alpha}^{\dagger} a_{m\beta} + a_{m\beta}^{\dagger} a_{m\alpha}) + \tfrac{1}{2}\sum_{i\alpha\beta} K_{i\alpha\beta} q_{i\alpha\beta}^{2} \qquad (1)$$

where $K_{i\alpha\beta} = M_{i\alpha\beta}\omega_{i\alpha\beta}^{2}$ is the stiffness, $G_{i\alpha\beta}$ is the electron-phonon coupling constant. We will further follow the procedure of an adiabatic exclusion for the phonon coordinate $q_{i\alpha\beta}$.

We solve for the Schrodinger equation $H\Psi = E\Psi$ by a linear combination of one-particle one-band states $\Psi = c_{\alpha} a_{\alpha}^{\dagger} |0> + c_{\beta} a_{\beta}^{\dagger} |0>$ (site index omitted) where $|0>$ is the vacuum state, $\alpha$ and $\beta$ are two narrow electronic bands ($\approx$ two levels). Inserting we get the averages:

$$<a_{\alpha}|H|a_{\alpha}^{\dagger}> = E_{\alpha} + \tfrac{1}{2}\sum_{i\alpha\beta} K_{i\alpha\beta} q_{i\alpha\beta}^{2}$$

$$<a_{\beta}|H|a_{\beta}^{\dagger}> = E_{\beta} + \tfrac{1}{2}\sum_{i\alpha\beta} K_{i\alpha\beta} q_{i\alpha\beta}^{2}$$

$$<a_{\beta}|H|a_{\alpha}^{\dagger}> = \sum_{i\alpha\beta} G_{i\alpha\beta} q_{i\alpha\beta} \qquad (2)$$

Anticipating further needs we also introduce the Jahn-Teller energies $E_{JTi\alpha\beta} = G_{i\alpha\beta}^{2}/2K_{i\alpha\beta}$ [7]. The obtained secular equation for the energy is

$$(<a_{\alpha}|H|a_{\alpha}^{\dagger}> - E)(<a_{\beta}|H|a_{\beta}^{\dagger}> - E) - <a_{\beta}|H|a_{\alpha}^{\dagger}><a_{\alpha}|H|a_{\beta}^{\dagger}> = 0$$

with two roots reading

$$E(q_{i\alpha\beta})_{\pm} = \tfrac{1}{2}\{(<a_{\alpha}|H|a_{\alpha}^{\dagger}> + <a_{\beta}|H|a_{\beta}^{\dagger}>) \pm$$

$$\sqrt{[(<a_{\alpha}|H|a_{\alpha}^{\dagger}> - <a_{\beta}|H|a_{\beta}^{\dagger}>)^{2} + <a_{\beta}|H|a_{\alpha}^{\dagger}><a_{\alpha}|H|a_{\beta}^{\dagger}>]\}} =$$

$$\tfrac{1}{2}\{\sum_{i\alpha\beta} K_{i\alpha\beta} q_{i\alpha\beta}^{2} + E_{\alpha} + E_{\beta} \pm \sqrt{[(2\sum_{i\alpha\beta} G_{i\alpha\beta} q_{i\alpha\beta})^{2} - (E_{gap\alpha\beta})^{2}]}\} \qquad (3)$$

where $E_{gap\alpha\beta} = |E_{\alpha} - E_{\beta}|$ is the interband (interlevel) energy gap.

Eqn. (3) gives the two-branch adiabatic electron energies forming a double-well dependence on the phonon coordinates below the original energy gap and a single-well dependence above the gap. (See Figure 1 for an example.) The positions of the wells along the coupled phonon coordinate $q_{i\alpha\beta}$ axis are obtained by minimizing the adiabatic energy (3) to read

$$q_{i\alpha\beta0} = \pm\sqrt{(2E_{JTi\alpha\beta}/K_{i\alpha\beta})}\sqrt{[1 - (E_{gap\alpha\beta}/4E_{JTi\alpha\beta})^2]} \qquad (4)$$

The coupled-phonon (vibronic) Hamiltonian is ($\eta=h/2\pi$):

$$H_{vib} = \sum_{m\alpha}E_{m\alpha}a_{m\alpha}^{\dagger}a_{m\alpha} + \sum_{m\alpha\beta}G_{m\alpha\beta}q_{m\alpha\beta}(a_{m\alpha}^{\dagger}a_{m\beta}+a_{m\beta}^{\dagger}a_{m\alpha}) +$$

$$½\sum_{m\alpha\beta}K_{m\alpha\beta}q_{m\alpha\beta}^2 + ½(\eta^2/M)\sum_{m\alpha\beta}(\partial^2/\partial q_{m\alpha\beta}^2) \qquad (5)$$

This is the Hamiltonian of a displaced harmonic oscillator vibrating in a double well potential [7]. The bottom positions of the wells are located at $\pm q_{\alpha\beta0}$, as given by eqn. (4). They correspond to the lower symmetries to which the unstable central configuration has collapsed. The exact solution for the double-well oscillator not being available, we solve for it by a linear combination of the eigenstates for the left-hand and right-hand wells. The eigenstates overlap as $<\chi(x+q_0)|\chi(x-q_0)> = \exp(-\xi_{0\alpha\beta}^2)$ to give a tunneling splitting of magnitude [7]

$$\Delta\tau_{sp} = E_{gap\alpha\beta}\exp(-\xi_{\alpha\beta0}^2),$$

$$\xi_{\alpha\beta0} = \sqrt{(M_{\alpha\beta}\omega_{\alpha\beta}^2/\eta\omega)}q_{\alpha\beta0}, \quad q_{i\alpha\beta0} = -(G_{i\alpha\beta}/K_{i\alpha\beta})(a_{m\alpha}^{\dagger}a_{m\beta} + a_{m\beta}^{\dagger}a_{m\alpha}) \qquad (6)$$

is the mode coordinate in dimensionless units. We note that the above reasoning is holding good for $4E_{JT} \gg E_{gap}$ which is the small-polaron condition leading to $\Delta\tau_{sp} \ll E_{gap}$.

The large-polaron condition is $4E_{JT} \geq E_{gap}$. The tunneling splitting is still given as above but in this case care should be taken in so far as the vibronic level is closer to the barrier top. Now the exponent in (6) is small, so expanding in a power series $\exp(-\xi_{\alpha\beta0}^2) \approx 1 - \xi_{\alpha\beta0}^2$ we get

$$\Delta\tau_{lp} \approx E_{gap\alpha\beta}(1 - \xi_{\alpha\beta0}^2) = E_{gap\alpha\beta}[1 - \sqrt{(M_{\alpha\beta}\omega_{\alpha\beta}/\eta)}(G_{\alpha\beta}/K_{\alpha\beta})(a_{m\alpha}^{\dagger}a_{m\beta} + a_{m\beta}^{\dagger}a_{m\alpha})] \quad (7)$$

Now $\Delta\tau_{lp} \leq E_{gap}$.

In the case of an intermediate coupling, the tunneling splitting reads likewise:

$$\Delta\tau = E_{gap}\exp(-\xi_0^2) \qquad (8)$$

which means that an energy gap $E_{gap}$ is reduced exponentially into a tunneling splitting $\Delta\tau$ through the polaron tunneling transform. We stress that while the original gap $E_{gap}$ is characteristic of a decoupled two-level system, the polaron-transform tunneling splitting (or tunneling gap) $\Delta\tau$ is inherent for the phonon-coupled two-level system.

### 3. Conceivable links with nuclear systems

In so far as the electron-phonon mixing Hamiltonian (1) is one of the second-quantization forms for the coupling of fermion to boson fields (electron to phonon fields in solid state), we are tempting to speculate on some obvious implications. Other examples of fermion-to-boson coupling are provided by quantum electrodynamics (electron to photon fields), as well as nuclear forces (nucleon to π-meson fields) [8]. On the other hand vibronic effects in chemistry and solid state physics are described at length in a number of good monographs [9-11].

A profound conclusion of the first example (present one) is the breaking down of inversion symmetry and the related appearance of an inversion electric dipole which gives rise to a number of important implications, such as the occurrence of vibronic VdW electrostatic interactions of immense magnitude in dilute systems based on the enhanced polarizability of phonon-coupled two-level systems. We stress on *electrostatic*, since a similar theory can be devised of *magnetostatic* coupling leading to the magnetic analogue.

In reference to nuclear matter, there may be an analogical symmetry lowering leading to the appearance of the oblate and prolate nuclei as the benchmark of a nuclear vibronic interaction. We remind of other benchmark cases, such as the left- and right-handed neutrinos resulting from parity non-conservation. All these suggest close interdisciplinary links which lead to the symmetry breaking as a most fundamental law of nature. We plan to address the problem in greater detail elsewhere.

Now by differentiating the adiabatic energy of off-center defects [11]:

$$E_L(\{Q_i\}) = \tfrac{1}{2}\{\sum_i K_i Q_i^2 - [\sum_i (2G_i Q_i)^2 + E_{gap\alpha\beta}^2]^{1/2}\} \qquad (9)$$

with respect to $Q_i$ we arrive at the equation for the extremal surface:

$$Q_x^2/Q_{x0}^2 + Q_y^2/Q_{y0}^2 + Q_z^2/Q_{z0}^2 = 1 \qquad (10)$$

with semiaxes reading

$$Q_{x0} = [(2E_{JTx}/K_x)(1-\eta_x^2)]^{1/2}, \quad \eta_x = E_{\alpha\beta}/4E_{JTx}$$

$$Q_{y0} = [(2E_{JTy}/K_y)(1-\eta_y^2)]^{1/2}, \quad \eta_y = E_{\alpha\beta}/4E_{JTy}$$

$$Q_{z0} = [(2E_{JTz}/K_z)(1-\eta_z^2)]^{1/2}, \quad \eta_z = E_{\alpha\beta}/4E_{JTz} \qquad (11)$$

where $E_{JT}$ are the Jahn-Teller energies, $E_{\alpha\beta}$ are the energy gaps.

For small vibronic polarons $\eta_x, \eta_y, \eta_z \ll 1$ and the extremal semiaxes are simply related to the observed semiaxes, namely,

$$Q_{x0} = \sqrt{(2E_{JTx}/K_x)}$$

$$Q_{y0} = \sqrt{(2E_{JTy}/K_y)}$$

$$Q_{z0} = \sqrt{(2E_{JTz}/K_z)} \qquad (12)$$

which become easier to operate with. For a large vibronic polaron, $\eta_x, \eta_y, \eta_z \leq 1$, then $1-\eta_x^2 \ll 1$, etc. so that one must use the complete eqn. (11) The semi-axes are small, by virtue of the $\eta_x$, etc. containing factors.

Now, if the Jahn-Teller energies and stiffness factors along the coordinate axes are all but equal to each other: $Q_{x0} \approx \sqrt{(2E_{JT}/K)}$, etc. then the nuclear shape is a sphere. The deformed spheroid shapes are expressed likewise: The oblate (pancake) shape obtains for $Q_{z0} < Q_{x0} =$

$Q_{y0}$, while the prolate (cigar) shape arises for $Q_{z0} > Q_{x0} = Q_{y0}$. Both are traditionally explained by a nonuniform distribution of extra nucleonic density over a closed shell nucleus.

Indeed, for a closed shell nucleus the j-shells of single-particle orbitals are either fully vacant or fully occupied and the nucleonic density distribution is spherical [2]. On adding an extra particle to the first empty j-shell, its density distribution will be concentrated in the equatorial plane. This generates a non-spherical field which aligns other in-coming particles to fit their orbital planes to the equator (aligned coupling). The resulting nucleonic density is deformed and corresponds to an oblate spheroid. Alternatively, on putting the extra nucleon in an orbital with density distribution concentrated along the polar axis, the deformed form will correspond to the prolate spheroid.

We propose an alternative explanation in terms of the quantum-mechanical off-center effect. To see where we are note that $2E_{JT}/K = (G/K)^2$ which leads to $Q_0 = G/K \sim 1$ fermi = $10^{-5}$ Å for a small polaron. It follows that $E_{JT} \sim 5 \times 10^{-11}$ K, inconsistent with a small-polaron status. This suggests that the solution has to be sought using the complete equation (11).

For this reason we go back to eqns. (11) and set $\eta_x = \eta_y = \eta_z$, $E_{JTx} = E_{JTy} = E_{JTz}$, and also $K_x = K_y = K_z$, $G_x = G_y = G_z$ to describe a sphere. From eqns. (12) we now get a quadratic equation for $E_{JT}$ (coordinate index dropped):

$$(2/K) E_{JT}^2 - Q_0^2 E_{JT} - (2/K)(E_{gap}/4)^2 = 0 \tag{13}$$

with roots reading

$$E_{JT\pm} = (K/4)\{Q_0^2 \pm \sqrt{[Q_0^4 + (E_{gap}/K)^2]}\} \tag{14}$$

where $Q_0 = r_0 A^{1/3}$ is the spherical-nucleus radius, $r_0 = 1.1$ fermi, A is the atomic weight, the formula holding good for all but the very light nuclei [10]. For an estimate of the spherical case, we set $E_{gap} = 1$ keV, $K = 2.5 \times 10^{-2}$ keV/fermi$^2$. From eqn, (14) we now get $E_{JT} = 13$ keV, $E_{gap}/4E_{JT} = 0.02$, well below the mode-softening limit $E_{gap}/4E_{JT} = 1$. From these data we calculate a coupling constant $G = \sqrt{(2E_{JT}K)} = 0.8$ keV/ fermi. We also obtain $Q_0 = 5.2$ fermi for $Pd^{106}$ and deduce $h\nu = 0.5$ MeV from Fig.2.3 of Ref. [5]. The resulting double-well potential is shown in Figure 1. The analysis of a spheroid shape may be carried out along similar lines.

### 4. Comments

It is remarkable, though not at all strange, what the above crude estimate shows in that the small-polaron condition makes way for the materialization of vibronic effects in nuclear matter. This is because the small-polaron case often leads to localization overpopulating the respective off-center sites at the expense of the other sites. This may also give preference to cigar-shaped distortions over oblate-shaped distortions. The large-polaron condition most often provides for unimpeded though still hindered rotations around the central lattice site.

Interionic sites for off-center displacements are those of minimum potential energy (potential wells) at some distance from the center, called *off-center displacement*. Similar minimum energy sites may also be found in amorphous solids and why not in nuclear matter too. The extremal sites given, the off-center displacements are a very likely possibility.

The quantum-mechanical off-center effect has first been introduced as regards observed off-site ions in crystalline solids [12], though vibronic models have later been considered relative to amorphous solids too [13].

Herein, we have suggested an alternative explanation for the deformations of nuclei by means of that same off-center (off-site, off-axis, site-splitting) effect based on the breaking of site symmetry by vibronic effects in solid state physics. Interdisciplinary studies have often been carried out between solid state physics and chemistry or nuclear physics, and vice versa, yielding useful results. For examples one could point to the nuclear shell model, the quantum-mechanical oscillator model, etc. to mention a few. The practice of borrowing methods and techniques from one discipline to another has proved fruitful, though it may also provide a deeper insight into a common phenomenon. Presently it is the breaking of a higher symmetry through the mixing of fermionic states as they couple to a boson field. This comes as nucleon-to-meson coupling in nuclear matter and electron-to-phonon coupling in solids.

One could also extend the analogy further and mention the appearance of electromagnetic off-axis effects occurring in light propagation which result from fermion-to-photon mixing. All these may have a common origin in the appearance of an universal symmetry-breaking law.

The present study is aimed at drawing attention. A complete second quanrization will further have to be applied for a thorough description of off-center effects within nuclei. Identifying the nature of the symmetry-breaking vibrations is an immense necessity to satisfy the critics.

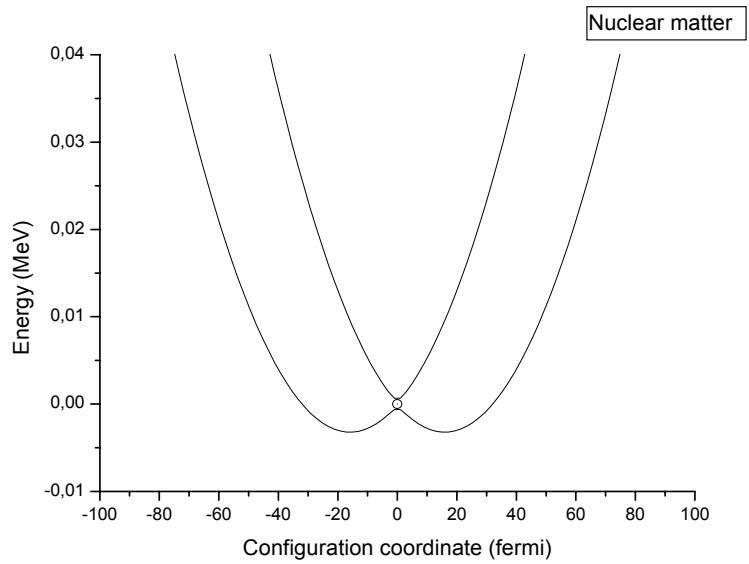

Figure 1: Illustrative adiabatic potentials for 1D off-center effects in a sample nuclear matter following eqn. (9). The parameters used for the calculations are: fermion-boson coupling constant G = 0.8 keV/Fermi, stiffness K = 0.025 keV/Fermi, fermion energy gap $E_{gap}$ = 1 keV. The interwell barrier is 13 keV.